\documentclass[twocolumn,showpacs]{revtex4}
\usepackage{graphics}
\usepackage{epsfig}
\usepackage{graphicx}
\usepackage{epstopdf}
\usepackage{bm}

\begin{document}
\title{Time-resolved measurement of Landau-Zener tunneling in periodic potentials}
\author{A. Zenesini$^{1,2}$, H. Lignier${^1}$, G. Tayebirad${^3}$, J. Radogostowicz$^{1,2}$, D. Ciampini$^{1,2}$, R. Mannella$^{1,2}$, S. Wimberger${^3}$, O. Morsch${^1}$, E. Arimondo$^{1,2}$}
\affiliation{${^1}$ CNR-INFM and Dipartimento di Fisica `E.
Fermi', Largo Pontecorvo 3, 56127 Pisa, Italy\\ ${^2}$ CNISM, Unit\`{a} di Pisa, Largo Pontecorvo 3, 56127 Pisa, Italy \\ ${^3}$Institut
f\"{u}r theoretische Physik, Universit\"{a}t Heidelberg, D–69120,
Heidelberg, Germany}

\begin{abstract}
We report time-resolved measurements of Landau-Zener tunneling of
Bose-Einstein condensates in accelerated optical lattices, clearly
resolving the step-like time dependence of the band populations.
Using different experimental protocols we were able to measure the
tunneling probability both in the adiabatic and in the diabatic
bases of the system. We also experimentally determine the
contribution of the momentum width of the Bose condensates to the
width of the tunneling steps and discuss the implications for
measuring the jump time in the Landau-Zener problem.
\end{abstract}


\maketitle

Tunneling is one of the most striking manifestations of quantum
behaviour and has been the subject of intense research both in
fundamental and applied physics~\cite{razavy_03}. While tunneling
{\it probabilities} can be calculated accurately even for complex
quantum systems and have an intuitive interpretation as
statistical mean values of experimental outcomes, the concept of
tunneling {\it time} and its computation are still the subject of
debate even for simple systems~\cite{schulman_08}. How much time a
quantum system spends in a classically forbidden area, e.g.,
inside a barrier separating two potential wells, has been
theoretically calculated for a wide range of physical
systems~\cite{hauge_89,landauer_94} and measured in recent
experiments~\cite{eckle_08}. A well-known example of a simple
tunneling problem is Landau-Zener (LZ) tunneling, in which a
quantum system tunnels across an energy gap at an avoided crossing
of the system's energy levels. In recent years, LZ tunneling has
been used as a paradigm for studies of the the quantum Zeno effect
(controlling the decay by repeated measurements of the
systems~\cite{facchi_08}), and more generally in the observation
of deviations from a purely exponential decay of a quantum system
consisting of just a few energy levels. An example of the latter
was investigated by Raizen and
co-workers~\cite{wilkinson_97,niu_98,fischer_99}, where the decay
in a Wannier-Stark system was shown to deviate from an exponential
law at short times. Moreover, LZ tunneling provides a building
block for the quantum control~\cite{haroche_review} of complex
many-body systems~\cite{santoro_02}.
\begin{figure}[ht]
\includegraphics[width=7.5cm]{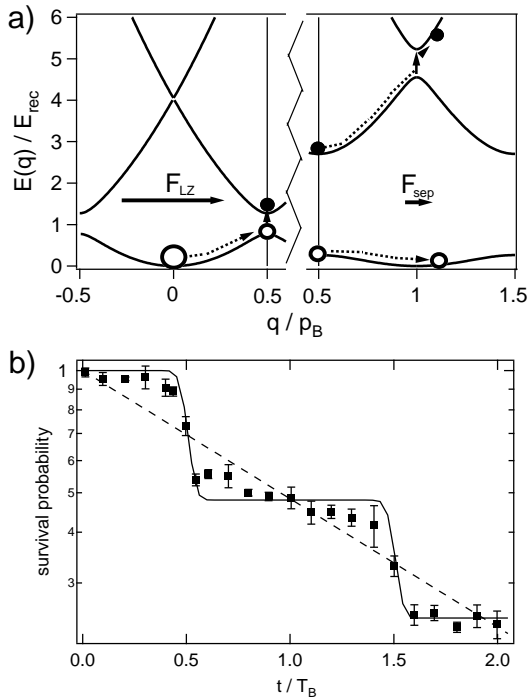}
\caption{\label{figure1} Time-resolved measurement of LZ
tunneling. (a) Experimental protocol (shown in the band-structure
representation of energy $E(q)$ versus lattice quasimomentum $q$).
{\it Left}: The lattice is accelerated, (partial) tunneling
occurs. {\it Right}: The acceleration is then suddenly reduced and
the lattice depth increased so as to `freeze' the instantaneous
populations in the lowest two bands; finally, further acceleration
is used to separate these populations in momentum space. (b)
Experimental results for $V_0=1\,\mathrm{E_{rec}}$ and
$F_0=0.383$, giving $T_B=0.826\,\mathrm{ms}$. The solid and dashed
lines are a numerical simulation of our experimental protocol and
an exponential decay curve for our system's parameters,
respectively.}
\end{figure}

According to the adiabatic theorem an infinitely slow change in
the Hamiltonian of a quantum system leads to an adiabatic
following of the instantaneous eigenstate, whereas a controlled
time dependence can cause transitions to other eigenstates. In the
case of a linear time dependence in a Hamiltonian describing two
crossing eigenstates that are repelled by a coupling energy, the
probability of a transition between these eigenstates is described
by the celebrated Landau-Zener model~\cite{landau_32,zener_32}.
While in most treatments of LZ-tunneling only the (asymptotic)
tunneling probability is considered, the full time dependence of
the LZ dynamics as determined by the time evolution of the
quantum-mechanical wavefunction also plays an important role,
especially if the asymptotic limit assumed in LZ theory is not
reached in a single transition. The characteristic time required
for the tunneling between initial and final quantum states to be
completed~\cite{mullen_89,landauer_94,vitanov_99} is sometimes
referred to as the "jump time".

In this Letter, we present experimental results, backed up by
numerical simulations, on the LZ dynamics and on the jump time for
a Wannier-Stark system realized with ultracold atoms forming a
Bose-Einstein condensate (BEC) inside an optical
lattice~\cite{anderson_98,morsch_01}. This system gives us an
optimal control of single as well as sequences~\cite{sias_07} of
LZ processes. In contrast to the above mentioned experiments with
cold atoms~\cite{wilkinson_97,fischer_99} our BEC has an initial
width in momentum space that is much smaller than the
characteristic momentum scale of the problem given by the momentum
width $p_B = 2p_{rec} = 2\pi \hbar  /d_L$ of the first Brillouin
zone of a periodic potential with lattice constant $d_L$. This
possibility of engineering the initial state for the LZ process in
momentum space enables us to observe the time dependent dynamics
with a non-exponential decay in the survival probability for
single or multiple LZ crossings~\cite{note} as well as
oscillations in the transition probability after the system has
passed the avoided crossing (as predicted for LZ tunneling in
atomic Rydberg states~\cite{rubbmark_81} and experimentally
observed in a wave-optical two-level
system~\cite{bouwmeester_95}), the only limitation being the
initial momentum width of the condensates and nonlinear effects.
This possibility of engineering the initial state for the LZ
process in momentum space enables us to observe the time dependent
dynamics with a non-exponential decay in the survival probability
for single or multiple LZ crossings~\cite{note} as well as
oscillations in the transition probability after the system has
passed the avoided crossing,  the only limitation being the
initial momentum width of the condensates and nonlinear effects.
Such oscillations were  predicted for LZ tunneling in atomic
Rydberg states~\cite{rubbmark_81} and experimentally observed in a
wave-optical two-level system~\cite{bouwmeester_95}). Our
experiments are similar to recent studies of LZ transitions at
avoided crossings in the energy levels of a solid-state artificial
atom~\cite{berns_08}, but the high level of control over the
light-induced periodic potential also allowed us to measure the
tunneling dynamics in different eigenbases of the system's
Hamiltonian.

In our experiments we created Bose-Einstein condensates of
$5\times 10^4$ rubidium-87 atoms inside an optical dipole trap
(mean trap frequency around $80\,\mathrm{Hz}$). A one-dimensional
optical lattice created by two counter-propagating, linearly
polarized gaussian beams was then superposed on the BEC by ramping
up the power in the lattice beams in $100\,\mathrm{ms}$. The
wavelength of the lattice beams was $\lambda=842\,\mathrm{nm}$,
leading to a sinusoidal potential with lattice constant
$d_L=\lambda/2=421\,\mathrm{nm}$. A small frequency offset between
the two beams could be introduced through the acousto-optic
modulators in the setup, allowing us to accelerate the lattice in
a controlled fashion.

The time-resolved measurement of LZ tunneling was then effected as
follows [see Fig. 1 (a)]. After loading the BEC into the ground
state energy band of an optical lattice of depth $V_0$ as
described above, the lattice was accelerated with acceleration
$a_\mathrm{LZ}$ for a time $t_\mathrm{LZ}$ by chirping the
frequency offset between the lattice beams, resulting in a
corresponding force $F_\mathrm{LZ}$ on the atoms in the rest frame
of the lattice. The lattice thus acquired a final velocity
$v=a_\mathrm{LZ} t_\mathrm{LZ}$. During $t_\mathrm{LZ}$ the
quasimomentum of the BEC swept the Brillouin zone, and at
multiples of half the Bloch time $T_\mathrm{B}=2 \pi
\hbar(Ma_\mathrm{LZ} d_{\mathrm L})^{-1}$ (where $M$ is the atomic
mass) {\it i.e.} at times $t=(n+1/2)T_\mathrm{B}$ ($n=0,1,2,...)$
when the system was close to the Brillouin zone edge, tunneling to
the upper band became increasingly likely. At time $t=t_{\mathrm
LZ}$ the acceleration was abruptly reduced to a smaller value
$a_\mathrm{sep}$ and the lattice depth was increased to
$V_\mathrm{sep}$ in a time $t_\mathrm{ramp}\ll T_\mathrm{B}$.
These values were chosen in such a way that at time
$t=t_\mathrm{LZ}$ the probability for Landau-Zener tunneling from
the lowest to the first excited energy band dropped from between
$\approx 0.1-0.9$ (depending on the initial parameters chosen) to
less than $\approx 0.01$, while the tunneling probability from the
first excited to the second excited band remained high at about
$0.95$. This meant that at $t=t_\mathrm{LZ}$ the tunneling process
was effectively interrupted and for $t>t_\mathrm{LZ}$ the measured
survival probability $P(t)=N_0/N_\mathrm{tot}$ (calculated from
the number of atoms $N_0$ in the lowest band and the total number
of atoms in the condensate $N_\mathrm{tot}$) reflected the
instantaneous value $P(t=t_\mathrm{LZ})$.

The lattice was then further accelerated for a time
$t_\mathrm{sep}$ such that $a_\mathrm{sep}t_\mathrm{sep}\approx 2n
p_\mathrm{rec}/M$ (where typically $n=2$ or $3$). In this way,
atoms in the lowest band were accelerated to a final velocity
$v\approx 2n p_\mathrm{rec}/M$, while atoms that had undergone
tunneling to the first excited band before $t=t_\mathrm{LZ}$
underwent further tunneling to higher bands with a probability
$>0.95$ and were, therefore, no longer accelerated. At time
$t_\mathrm{sep}$ the lattice and dipole trap beams were suddenly
switched off and the expanded atomic cloud was imaged after
$23\,\mathrm{ms}$. In these time-of-flight images the two velocity
classes $0$ and $2n p_\mathrm{rec}/M$ were well separated and the
atom numbers $N_0$ and $N_\mathrm{tot}$ could be measured
directly. Since the populations were "frozen" inside the energy
bands of the lattice, which represent the adiabatic eigenstates of
the total Hamiltonian of the system, this experiment effectively
measured the time dependence of the LZ survival probability $P_a$
in the {\it adiabatic} basis. The result of a typical measurement
is shown in Fig.~1 (b). One clearly sees two "steps" at times
$t=0.5\,T_B$ and $t=1.5\,T_B$, which correspond to the instants at
which the atoms cross the Brillouin zone edges, where the lowest
and first excited energy bands exhibit avoided crossings. For
comparison, the result of a numerical simulation (integrating the
linear Schr\"{o}dinger equation for the experimental protocol) as
well as an exponential decay as predicted by LZ theory are also
shown.

\begin{figure}[ht]
\includegraphics[width=7cm]{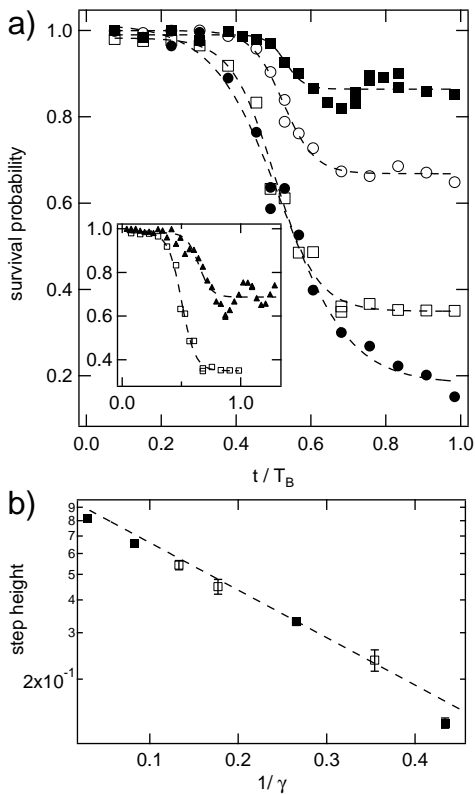}
\caption{\label{figure1} (a) LZ survival probability in the
adiabatic basis for a fixed force $F_0=1.197$ and different
lattice depths (filled squares: $V_0=2.3\,\mathrm{E_{rec}}$; open
circles: $V_0=1.8\,\mathrm{E_{rec}}$; open squares:
$V_0=1\,\mathrm{E_{rec}}$; filled circles:
$V_0=0.6\,\mathrm{E_{rec}}$). The dashed lines are sigmoid fits to
the experimental data. {\it Inset}: Survival probability in both
the adiabatic (open squares) and diabatic (filled triangles) bases
for $V_0=1\,\mathrm{E_{rec}}$ and $F_0=1.197$. (b) Step height $h$
as a function of the inverse adiabaticity parameter $1/\gamma$ for
varying lattice depth and $F_0=1.197$ (open symbols), and for varying
force with fixed $V_0=1.8\,\mathrm{E_{rec}}$ (filled symbols). The dashed
line is the prediction of Eq.~(3) for the LZ
tunneling probability.}
\end{figure}

The LZ tunneling probability can be calculated by considering a
two-level system with the adiabatic Hamiltonian
\begin{equation}
H_a=H_d+V=\alpha t \sigma_y+\frac{\Delta E}{2}
\sigma_x, \label{hamiltonian}
\end{equation}
where $\sigma_i$ are the Pauli matrices. The eigenstates
of the diabatic Hamiltonian $H_d$, whose eigenenergies
vary linearly in time, are mixed by the potential $V$
characterized by the energy gap $\Delta E$. Applying the Zener
model~\cite{zener_32} to our case of a BEC crossing the Brillouin
zone edge leads to a band gap  $\Delta E= V_0/2$ and to $\alpha=2v_{\mathrm rec} M a_{\mathrm LZ}=2F_0
E_{\mathrm rec}^2/(\pi \hbar)$, with  $E_{\mathrm rec}=\hbar^2 \pi^2/(2 M
d_{\mathrm l}^2)$ the recoil energy and   $F_0 = M a_{\mathrm LZ} d_{\mathrm
L}/E_{\mathrm rec}$ the
dimensionless force. The limiting value of the adiabatic and
diabatic Lanadau-Zener survival probabilities (for $t$ going from $-\infty$ to
$+\infty$) in the eigenstates of $H_a$ and $H_d$,
respectively, is
\begin{equation}
P_a(t\rightarrow+\infty) =
1-P_d(t\rightarrow+\infty)=1-P_{\mathrm LZ},
\label{probability}
\end{equation}
where \begin{equation} P_{\mathrm
LZ}=e^{-\frac{\pi}{\gamma}}
\label{LZresult}
\end{equation} with the adiabaticity
parameter $\gamma=4\hbar\alpha(\Delta E)^{-2}$
is the standard LZ tunneling probability~\cite{note2}.

In order to test whether this tunneling probability correctly
predicts the height of the step corresponding to a single LZ
tunneling event, we performed the experiment described above for a
variety of values of $V_0$ and $F$. Figure 2(a) shows the first
tunneling step for different lattice depths $V_0$, measured in
units of $E_{\mathrm rec}$ at a given LZ acceleration determined
by  $F_0 $. The steps can be well fitted with a sigmoid function
\begin{equation}
P_a(t) = 1-\frac{h}{1+\exp((t_0-t)/\Delta
t_{\mathrm{LZ}})},
\end{equation}
where $t_0$ is the position of the step (which can deviate
slightly from the expected value of $0.5\,T_B$, e.g. due to a
non-zero initial momentum of the condensate), $h$ is the step
height given by Eq. (\ref{LZresult}) and $\Delta t_{\mathrm{LZ}}$
represents the width of the step. As expected, the step height $h$
follows the LZ result for the tunneling probability (see
Fig.~2(b)).

We note here that while the experimental protocol described above
measures the LZ tunneling probability in the {\it adiabatic
basis}, it is also possible to make analogous measurements in the
{\it diabatic basis} of the unperturbed free-particle
wavefunctions (plane waves with a quadratic energy-momentum
dispersion relation) by abruptly switching off the lattice and the
dipole trap after the first acceleration step. In this case, after
a time-of-flight the number of atoms in the $v=0$ and
$v=2p_\mathrm{rec}/M$ momentum classes are measured and from these
the survival probability (corresponding to the atoms remaining in
the $v=0$ velocity class relative to the total atom number) is
calculated. The inset of Fig. 2(a) (filled triangles) shows such a
measurement. Again, a step around $t=0.5\,T_B$ is clearly seen, as
well as strong oscillations for $t>0.5\,T_B$. While (weaker)
oscillations can also be measured in the adiabatic basis (see the
results for $V_0=2.3\,\mathrm{E_{rec}}$ in Fig. 2(a)), they are
much stronger and visible for a wider range of parameters in the
diabatic basis, as expected from theoretical
calculations~\cite{vitanov_99}.

The widths $\Delta t_{\mathrm{LZ}}$ corresponding to the steps
shown in Fig. 2(a) should, according to our interpretation,
reflect the "tunneling time" or "jump time" for LZ tunneling
$\Delta t_\mathrm{LZ}=\Delta v_\mathrm{LZ}/a_\mathrm{LZ}$ during
which the probability of finding the atoms in the lowest energy
band goes from $P_a(t=0)=1$ to its asymptotic LZ
value $P_\mathrm{LZ}$. In the theoretical literature
this tunneling time has been calculated by several
authors~\cite{mullen_89,landauer_94,vitanov_99}.
Vitanov~\cite{vitanov_99} defines the jump time in the adiabatic
basis as
\begin{equation}
\tau^{\mathrm jump}_a=\frac{P_a(t=+\infty)}{P'_a(t=t_0)},
\label{jumptime}
\end{equation} where $P'_a(t=t_0)$ denotes the time derivative of the tunneling probability $P_a(t)$ evaluated at the crossing point of $H_{\mathrm
a}$. A sigmoidal function for $P_a(t)$ leads to $\tau^{\mathrm
jump}_a=4\Delta t_\mathrm{LZ}$. For large values of $\gamma$,
which is the regime of our experiments, the theoretical jump time
is given by $\tau^{\mathrm jump}_a/T_B\approx
(V_0/E_\mathrm{rec})/8$. From our sigmoidal fits we retrieve
$\tau^{\mathrm jump}_a/T_\mathrm{B}\approx 0.15-0.35$
(corresponding to absolute jump times between $50\,\mathrm{\mu s}$
and $200\,\mathrm{\mu s}$), whereas the theoretical values for our
experimental parameters are in the region of $0.1-0.15$. We
interpret this discrepancy as being due the fact that in our
experiment the condensate does not occupy one single quasimomentum
but is represented by a momentum distribution of a finite width
$\Delta p/p_B\gtrsim 0.1$ due to the finite number of lattice
sites (around $50$) it occupies and the effects of atom-atom
interactions. Also, during the acceleration process dynamical
instabilities can further broaden the momentum distribution.

In order to test the dependence of the measured step width on the
width of the momentum distribution of the condensate, we created
initial distributions of different widths using a dynamical
instability in a controlled way~\cite{cristiani_04}. The
condensate was loaded into a lattice moving at a finite velocity
corresponding to quasimomentum $q=-0.3\,p_B$ and held there  for a
time up to $3\,\mathrm{ms}$. During this time the dynamical
instability associated with the negative effective mass at
$q=-0.3\,p_B$ led to an increase in the momentum width of the
condensate. After this preparatory stage, the LZ dynamics was
measured as described above and the width of the tunneling step
was measured [see Fig.~3(a)]. As expected, the larger the initial
momentum width of the condensate, the larger the step width [Fig.
3(b)]. This was also confirmed by a numerical integration of the
Schr\"{o}dinger equation in which the momentum width of the
condensate was varied by changing the initial trap frequency. The
simulation also showed that for a vanishing momentum width, the
step width still remains finite and in that limit directly
reflects the jump time given by Eq. (\ref{jumptime}).

\begin{figure}[ht]
\includegraphics[width=6cm]{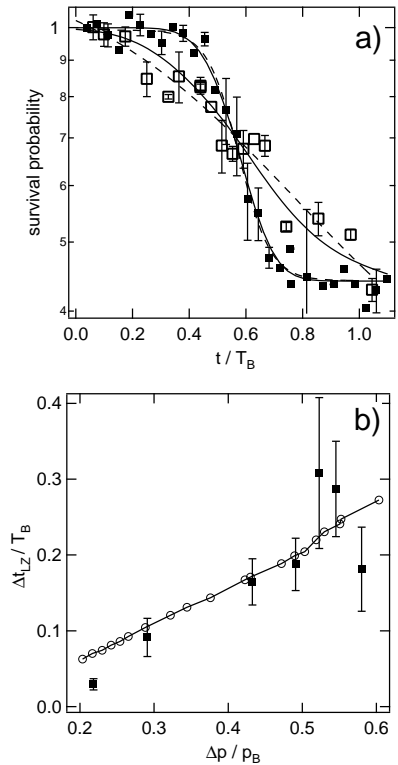}
\caption{\label{figure1} LZ transition for different
momentum widths of the condensate. (a) Survival probability for
$\Delta p/2p_{\mathrm{rec}}=0.2$ (filled squares) and $\Delta
p/2p_{\mathrm{rec}}=0.6$ (open squares). The solid and dashed
lines are the results of a numerical simulation and of a sigmoid
fit, respectively. (b) Step width $\Delta t_{LZ}$ of the LZ transition
as a function of the momentum width of the condensate. The solid
line is the result of a numerical simulation.}
\end{figure}

In summary, we have measured the full dynamics of a single LZ
transition of matter waves in an accelerated optical lattice. In
both the adiabatic and diabatic bases the step-like behaviour as
well as oscillations of the survival probability were clearly seen
and agree with theoretical predictions. While there is
quantitative agreement between the measured and theoretically
expected step heights, the jump times calculated from the step
widths are larger by $\approx 50-100\%$ than the theoretical
results because of the finite initial momentum widths of the
condensates in our experiments.

In future investigations one could reduce the initial momentum
width by using, e.g., appropriate trap geometries or by
controlling the nonlinearity through Feshbach resonances. This
would allow one to obtain a more quantitative measurement of the
jump time and enable a comparison with theoretical results related
to the minimum time for a single LZ crossing limited by
fundamental quantum (or wave, see~\cite{bouwmeester_95})
mechanical properties~\cite{caneva_09}. Also, clearer observations
of the short-time oscillations (happening on even shorter time
scales), whose signatures can be seen in Fig. 2(a) should be
possible in this way. Our method can also be used to study
multiple LZ crossings, e.g., in order to observe St\"{u}ckelberg
oscillations.

We gratefully acknowledge funding by the E.U. project "NAMEQUAM",
the CNISM "Progetto Innesco 2007" and the Excellence Initiative by
the German Research Foundation (DFG) through the Heidelberg
Graduate School of Fundamental Physics (grant number GSC 129/1)
and the Global Networks Mobility Measures. We thank Martin
Holthaus and Tilman Esslinger for discussions and comments on the
manuscript.

\bibliographystyle{apsrmp}

\end{document}